\documentclass[manuscript]{aastex}

\usepackage{hyperref}

\def\be{\begin{equation}}
\def\ee{\end{equation}}
\def\ba{\begin{array}}
\def\ea{\end{array}}
\def\bea{\begin{eqnarray}}
\def\eea{\end{eqnarray}}
\def\bi{\begin{itemize}}
\def\ei{\end{itemize}}

\shorttitle{Symmetry energy probed by Cas A cooling}
\shortauthors{Newton et al.}

\begin{document}


\title{The cooling of the Cassiopeia A neutron star as a probe of the nuclear symmetry energy and nuclear pasta}


\author{William G. Newton, Kyleah Murphy\altaffilmark{1}, Joshua Hooker and Bao-An Li}
\affil{Department of Physics and Astronomy, Texas A\&M University-Commerce, Commerce, Texas 75429-3011, USA}




\altaffiltext{1}{Umpqua Community College, Roseburg, Oregon, 97470, USA}

\begin{abstract}
X-ray observations of the neutron star in the Cas A supernova remnant over the past decade suggest the star is undergoing a rapid drop in surface temperature of $\approx$ 2-5.5\%. One explanation suggests the rapid cooling is triggered by the onset of neutron superfluidity in the core of the star, causing enhanced neutrino emission from neutron Cooper pair breaking and formation (PBF). Using consistent neutron star crust and core equations of state (EOSs) and compositions, we explore the sensitivity of this interpretation to the density dependence of the symmetry energy $L$ of the EOS used, and to the presence of enhanced neutrino cooling in the bubble phases of crustal ``nuclear pasta''. Modeling cooling over a conservative range of neutron star masses and envelope compositions, we find $L\lesssim70$ MeV, competitive with terrestrial experimental constraints and other astrophysical observations. For masses near the most likely mass of $M\gtrsim 1.65 M_{\odot}$, the constraint becomes more restrictive $35\lesssim L\lesssim 55$ MeV. The inclusion of the bubble cooling processes decreases the cooling rate of the star during the PBF phase, matching the observed rate only when $L\lesssim45$~MeV, taking all masses into consideration, corresponding to neutron star radii $\lesssim 11$km.
\end{abstract}



\section{Introduction}

In 2009, the thermal emission from the neutron star (NS) in the Cassiopeia A (Cas A) supernova remnant was fit using a carbon atmosphere model \citep{Ho2009} in order to obtain an emitting area consistent with canonical neutron star radii. The resulting average effective surface temperature was $\langle T_{\rm eff} \rangle \approx 2.1\times10$K. Subsequent analysis of \emph{Chandra} data taken over the previous 10 years indicated a rapid decrease in $T_{\rm eff}$ by $\approx$4\% \citep{Heinke2010}. A recent analysis of \emph{Chandra} data from all X-ray detectors and modes concluded a more uncertain range of a 2-5.5\% temperature decline, cautioning that a definitive measurement is difficult due to the surrounding bright and variable supernova remnant \citep{Elshamouty2013}. The most recent results from the ACIS-S detector (which gives the $\approx 4\%$ temperature decline between 2000 and 2009) are shown in Fig.~1 along with the best fit line, and two lines indicating best estimates for the shallowest ($\approx 2\%$) and steepest ($\approx 5.5\%$) declines. We take the age of Cas A NS (hereafter CANS) in 2005 to be $\tau_{\rm CANS} \approx 335$ yrs based on the estimated date of the supernova $\approx 1680 \pm 20$ yrs \citep{Fesen2006a}.

Within the minimal cooling paradigm (MCP), which excludes all fast neutrino ($\nu$)-emission processes such as direct Urca (DU) but includes superfluid effects \citep{Page2004}, the rapid cooling of the CANS is interpreted as the result of enhanced $\nu$-emission from neutron Cooper pair (CP) breaking and formation in the NS core (the ``PBF'' mechanism), providing evidence for stellar superfluidity \citep{Shternin2011,Page2011,Ho2013}. Other proposed models \citep{Blaschke2012,Sedrakian2013} involve medium modification to standard $\nu$-emission processes such as modified Urca (MU) and nucleon Bremsstrahlung, or a phase transition to quark matter. 

Neutrons in the NS core are expected to form CPs in the $^3P_2$ channel, while the protons form $^1S_0$ CPs. The pairing gaps and corresponding local critical temperatures $T_{\rm c}$ for the onset of superfluidity are strongly density dependent, and suffer significant theoretical uncertainty. The maximum value of the neutron $^3P_2$ critical temperature  $T_{\rm cn}^{\rm max}$ determines the age of the NS when the PBF cooling phase is entered, $\tau_{\rm PBF}$, and can be tuned so that the PBF cooling trajectory passes through the observed temperature of the CANS at an age of $\approx  335$ years. The core temperature at the onset of the PBF phase, $T_{\rm PBF}$, controls the subsequent cooling rate; a higher $T_{\rm PBF}$ leads to a steeper cooling trajectory. Proton superconductivity in the core inhibits the MU cooling process, leading to a higher $T_{\rm PBF}$; the width and magnitude of the $^1S_0$ proton pairing gap profile can thus be tuned to alter the slope of the resulting cooling curve in the PBF phase. \cite{Shternin2011,Page2011} find $T_{\rm cn}^{\rm max} \approx 5-9\times10^8$K and proton superconductivity throughout the whole core is required to fit the position and steepness of the observed cooling trajectory.

In the MCP, three other parameters affect the cooling trajectories of NSs \citep{Page2004}:  the mass of light elements in the envelope of the star $\Delta M_{\rm light}$, here parameterized as $\eta =  \log \Delta M_{\rm light}/M_{\odot}$ \citep{Yakovlev2011}, the mass of the star $M$ and the equation of state (EOS) of nuclear matter (NM). The thermal spectrum from the CANS can be fit using light element masses $-13 < \eta < -8$ and a NS mass of $\approx 1.25 - 2 M_{\odot}$ with a most likely value of $\approx 1.65 M_{\odot}$ \citep{Yakovlev2011}. The presence of more light elements (larger $\eta$) in the envelope increases the thermal conductivity there, increasing the observed surface temperature for a given temperature below the envelope \citep{Yakovlev2011}. \cite{Shternin2011,Page2011} used the APR EOS \citep{Akmal1998,Heiselberg1999}; however, the NM EOS is still quite uncertain.

Nuclear matter models are characterized by their behavior around nuclear saturation density $n_0 = 0.16$ baryons fm$^{-3}$, around which much of our nuclear experimental information is extracted. Denote the energy per particle of nuclear matter by $E(n,\delta)$, where $\delta = 1-2x$ is the isospin asymmetry, and $x$ is the proton fraction. $\delta = 0$ corresponds to symmetric nuclear matter (SNM), and $\delta = 1$ to pure neutron matter (PNM). We define the \emph{symmetry energy} $S(n)$ in the expansion about $\delta=0$: $E(n,\delta) = E_{\rm 0}(\chi) + S(n)\delta^2 + ...$ . $S(n)$ encodes the energy cost of decreasing the proton fraction of matter. Expanding $S(n)$ about $\chi=0$ where $\chi = \frac{n-n_{\rm 0}}{3n_{\rm 0}}$, we obtain $S(n) = J + L \chi + ...$ where $J$ and $L$ are the symmetry energy and its slope at $n_0$. $L$ determines the stiffness of the NS EOS around $n_0$ and correlates with NS radii \citep{Lattimer2001}, crust thickness \citep{Ducoin2011} and the extent of so-called ``nuclear pasta''  phases in the inner crust \citep{Oyamatsu2007}. Terrestrial constraints on $L$ from measurements of nuclear neutron skins, electric dipole polarizability, collective motion and the dynamics of heavy ion collisions \citep{Li2008,Tsang2012,Newton2013a,Lattimer2013,Danielewicz2013} suggest $30 \lesssim L\lesssim 80$ MeV, although larger values are not ruled out. \emph{Ab initio} calculations of PNM with well defined theoretical errors offer constraints on $J$ and $L$ (Fig. 2), and constraints on $S(n)$ from neutron star observations result in ranges of $L$ in broad agreement with experiment \citep{Ozel2010,Steiner2010,Steiner2012,Steiner2013,Gearheart2011,Sotani2013}. In this letter we show that we can extract a conservative constraint $L\lesssim70$MeV within the MCP using the CANS data, and even more stringent constraints with reasonable assumptions about the mass of the star.

At the base of the neutron star crust, matter is frustrated and it becomes energetically favorable for the nuclei there to form cylindrical, slab or cylindrical/spherical bubble shapes - ``nuclear pasta'' \citep{Ravenhall1983,Hashimoto1984}. Searching for observational signatures of the nuclear pasta phases is one quest of neutron star astrophysics \citep{Pons2013}. Two rapid $\nu$-emission processes have been postulated to operate in the bubble phases of nuclear pasta:  neutrino-antineutrino pair emission \citep{Leinson1993} and DU \citep{Gusakov2004}. We refer to these two mechanisms collectively as bubble cooling processes (BCPs). The neutrino luminosity from the BCPs are comparable: $L_{\rm \nu}^{BCP} \sim 10^{40} T_9^6$ where $T_9 = T_{\rm core}/10^9$K. Compared with the MU neutrino luminosity $L_{\rm \nu}^{MU} \sim 10^{40} T_9^8$, the BCP becomes competitive with MU cooling at temperatures below $10^9$K - i.e. at ages of order the CANS. We thus expect the temperature to be lower at ages $\gtrsim 300$ yrs with BCPs active, and thus the PBF cooling trajectory shallower. In this letter we show that with BCPs active, calculated cooling trajectories are only marginally consistent with observations, and only if the EOS is particularly soft: $L\lesssim 45$ MeV. 

Two caveats must be stated. The carbon atmosphere model is preferred \emph{solely} on the grounds that the resultant emitting area is consistent with neutron star radii. Other atmosphere compositions are not ruled out, and would result in changes to the inferred $T_{\rm eff}$ by up to a factor of 2, changing the inferred ranges of $L$. Secondly, the $^1S_0$ neutron and proton pairing gaps are quite model-dependent and might be significantly enhanced in the bottom layers of the crust compared to the model we use here. This would significantly suppress the BCPs and weaken the latter constraints on $L$.

\section{Model}

We calculate crust and core EOSs consistently using the Skyrme nuclear matter (NM) model. We choose the baseline Skyrme parameterization to be the SkIUFSU model \citep{Fattoyev2012a,Fattoyev2012b}, which shares the same saturation density nuclear matter properties as the relativistic mean field (RMF) IUFSU model \citep{Fattoyev2010}, has isovector NM parameters obtained by fitting to \emph{ab-initio} PNM calculations, and describes well the binding energies and charge radii of doubly magic nuclei \citep{Fattoyev2012a}. Two parameters in the Skyrme model can be adjusted to systematically vary the symmetry energy $J$ and its density slope $L$ at $n_0$ while leaving SNM properties unchanged \citep{Chen2009}. The constraints from PNM at low densities induce a correlation $J = 0.167 L + 23.33$ MeV. In this work we create EOSs characterized by $L=30-80$ MeV; the resulting PNM EOSs are shown for $L=30, 50, 70$ MeV in Fig.~2. These Skyrme NM models are then used to construct NS core EOSs (including compositions and nucleon effective masses) with the additional constraint that $M_{\rm max} > 2.0M_{\odot}$ \citep{Demorest2010,Antoniadis2013}, and consistent crust EOSs, compositions, and density ranges for the bubble phases of nuclear pasta using a liquid drop model \citep{Newton2013}. The resulting transition densities are very close to the `PNM' sequence in Figs~6 and~15 of \cite{Newton2013}. For a star of fixed mass, as $L$ increases, the stellar radius and crust thickness increases (see, e.g., Fig.~2 of \citep{Hooker2013}) and the fraction of the crust by mass composed of the bubble phases decreases from $\sim 1/6$ at $L=30$ MeV to zero at $L\approx70$ MeV \citep{Newton2013}.

We use the thermal envelope model \citep{Potekhin1997}, neutron and proton $^{1}S_{0}$ gaps \citep{Chen1993} (model CCDK in \cite{Page2004}), neutron $^{3}P_{2}$ gap, and PBF model \citep{Yakovlev1999,Kaminker1999} used in \cite{Page2011}. We use the publicly available code NSCOOL to perform the thermal evolution \url{http://www.astroscu.unam.mx/neutrones/NSCool/}. The neutrino emissivity for the BCPs is from \citet{Leinson1993}. We perform calculations at the limiting values of $\eta = -8$ and $\eta = -13$, masses of $M=1.25M_{\odot}, 1.4M_{\odot}$, $1.6M_{\odot}$ and $1.8M_{\odot}$ and for EOSs in the range $L=30 - 80$ MeV.

\section{Results}

Fig.~3 illustrates the impact of $L$, $M$, $\eta$ and the inclusion of BCPs on fitting the \emph{position} of the CANS data. Each plot shows cooling trajectories without and with the BCPs (solid and dashed lines respectively) and for the limiting $T_{\rm cn}^{\rm max}$ values of 0K (no $^{3}P_{2}$ neutron pairing) (upper trajectories) and $10^9$K (lower trajectories). We plot the inferred CANS effective surface temperature as seen by the observer $T_{\rm eff}^{\infty}$ - i.e. gravitationally redshifted from the surface temperature at the star $T_{\rm eff}^{\infty} = (1+z)^{-1} T_{\rm eff}$ where $z = (1 - 2GM/Rc^2)^{-1/2} - 1$, a factor which depends on $M$ and $L$ (the latter determining the radius $R$ for fixed $M$). Each pair of trajectories $T_{\rm cn}^{\rm max}$ = 0K, $10^9$K, forms a cooling window inside which the observed temperature must fall. BCPs narrow the cooling window from the higher temperature limit at ages $\sim \tau_{\rm CANS}$: $T_{\rm cn}^{\rm max}$ = 0K, the BCPs have a noticeable cooling effect which lowers $T_{\rm eff}^{\infty}$ while at $T_{\rm cn}^{\rm max} = 10^9$K, free neutrons in the bubble phases have already undergone the superfluid transition and thus the BCPs are suppressed; we thus see little effect for those trajectories. It is important to note that enhancement of $^1S_0$ neutron and proton gaps will also suppress BCPs, giving results closer to the ``no BCP'' cases presented here.

A higher $\Delta M_{\rm light}$ leads to higher $T_{\rm eff}^{\infty}$ for a given core temperature, as illustrated comparing $\eta =$ -8 and -13 in Figs~3a,b for $L=50$MeV, $M=1.25 M_{\odot}$; the cooling window is thus elevated relative to the observed $T_{\rm eff}^{\infty}$. As $M$ increases, the central stellar density increases and the fraction of the core in which the protons are superconducting decreases, making the MU process more efficient and the star cooler at $\tau_{\rm CANS}$. Decreasing $L$ decreases the radius, thus requiring a higher surface temperature to produce the same stellar luminosity. These trends are illustrated in Figs 3c-h.

If the measured CANS temperature falls within the theoretical cooling window for a given set of parameters, then one can find a value of $T_{\rm cn}^{\rm max}$ for which the cooling trajectory passes through the average measured temperature $\langle T_{\rm eff}^{\infty} \rangle$. Table I summarizes the ranges of $L$ for selected masses, $\eta$=-8 and -13 and with and without BCPs, for which the CANS data falls within the cooling window. Considering the full ranges of parameters, a constraint of $L\lesssim70$ MeV is extracted. Fitting of the thermal emission suggests that the mass is likely above $1.4M_{\odot}$, which gives a more restrictive constraint of $L\lesssim 60$ MeV. The ranges for $T_{\rm cn}^{\rm max}$ obtained with and without BCPs when all other parameters are varied are $5.1-5.7\times10^8$ and $5.6-9\times10^8$ respectively; the inclusion of BCPs leads to a more restrictive range.

Fig.~4 shows cooling windows for four sets of parameters ($L$(MeV), $M/M_{\odot}$, $\eta$) = (30, 1.4, -13) (Fig. 4a), (40, 1.6, -13) (Fig. 4b), (60, 1.4, -8) (Fig. 4c),  (50,1.8,-13) (Fig. 4d), as well as the curves corresponding to the value of $T_{\rm cn}^{\rm max}$ that best fits $\langle T_{\rm eff}^{\infty} \rangle$. The limiting cooling rates given in Fig.~1 are indicated by the two straight lines intersecting at $\langle T_{\rm eff}^{\infty} \rangle$. Calculated trajectories should have slopes between these two lines as they pass through the average temperature. Even the 2\% temperature decline is relatively rapid, favoring a relatively high core temperature at an age $\tau_{\rm PBF}$ and thus favoring smaller stars (smaller $L$), smaller masses $M$, a larger $\Delta M_{\rm light}$ (larger $\eta$), and disfavoring BCPs. Note in particular, with active BCPs the best fit cooling curve in the PBF phase is significantly less steep than without BCPs, and matches only the shallowest inferred cooling rate, and then only for the lowest values of $L$. As $L$ increases beyond $50-60$ MeV, depending on mass, the curves become too shallow to match the data even with no BCPs operating. The ranges of $L$ satisfying the \emph{slope} range of the cooling curve inferred from observation \emph{as well as} the average temperature are given in the second part of Table I.

\section{Discussion and conclusions}

Being agnostic about the mass of the neutron star in Cas A and the mass of the light element blanket within the ranges inferred from fitting the thermal spectrum under the assumption of a carbon atmosphere ($1.25 M_{\odot} < M < 1.8M_{\odot}$, $10^{-13} < \Delta M_{\rm light} < 10^{-8}$), theoretical cooling curves pass through the average inferred surface temperature if $L\lesssim 70$ MeV. For a mass $M=1.6 M_{\odot}$ (close to the most likely inferred mass of $1.65 M_{\odot}$), the range becomes slightly more restrictive $L\lesssim 65$ MeV. 

Requiring the inferred cooling \emph{rate} to be matched within its range of uncertainty, the constraint on $L$ tends to become more restrictive still. With BCPs inactive, $L\lesssim 70$ MeV ($35\lesssim L\lesssim 55$ MeV) for $1.25 M_{\odot} < M < 1.8M_{\odot}$ ($M=1.6 M_{\odot}$). With BCPs active, cooling curves become shallower and we obtain our most restrictive constraints $L\lesssim 45$ MeV ($35\lesssim L\lesssim 45$ MeV) for $1.25 M_{\odot} < M < 1.8M_{\odot}$ ($M=1.6M_{\odot}$). The latter constraints correspond to neutron star radii $\lesssim11$km for the EOSs used here.

Accepting the MCP cooling model and the accuracy of X-ray measurements and interpretation, we can conclude either: (i) efficient cooling mechanisms are active in the bubble phases of nuclear pasta and $L\lesssim45$ MeV, or (ii) efficient cooling in nuclear pasta is suppressed, and $L\lesssim 70$ MeV. Such suppression could occur if the high density tail of the neutron $^1S_0$ pairing gap profile or the low density tails of the neutron $^3P_2$ or proton $^1S_0$ pairing gap profiles enhanced superfluidity in the bubble phases. Additionally, there might be other unexplored medium effects that inhibit the BCPs such as entrainment of crustal neutrons \citep{Chamel2012}. However, even at their most conservative, these constraints are competitive with experimental constraints $L\approx$ 30-80 MeV.

Together, the physics of many aspects of a neutron star surface and interior affect its temperature evolution; in this letter we have systematically examined the effect of two such aspects, namely the slope of the symmetry energy $L$ and the presence of enhanced cooling in the bubble phases while controlling for the behavior of other physical aspects. We must caution that we have not accounted for every possible parameter and variation thereof. We cannot rule out atmosphere models other than the carbon composition model upon which the current $\langle T_{\rm eff}^{\infty} \rangle$ is based; use of other models would shift the inferred range of $L$. Broadening the range of the $^1S_0$ proton pairing gap would inhibit MU cooling even more: this would raise the temperature at the onset of the PBF phase, steepening the cooling curve. Additional variations in the high density EOS could also shift the inferred range of $L$. As an example, \cite{Shternin2011} find 1.8$M_{\odot}$ stellar models that match the CANS cooling rate, whereas we do not. The APR EOS used there gives a maximum mass $M_{\rm max} \approx 1.9M_{\odot}$, below the observed lower limit, so their high mass models will tend to be more compact and allow steeper cooling trajectories. Stiffening the high density EOS to increase $M_{\rm max}$ above $\approx 2M_{\odot}$ will tend to decrease the cooling rate. Additionally, the crust model is not consistent with their core EOS, and although their gap profiles reach similar magnitudes as those used here, the gap profiles are different.

Despite these limitations, we have demonstrated that current cooling observations of the Cas A NS have the potential to impose strong constraints on the slope of the symmetry energy $L$ at saturation density and demonstrated for the first time that enhanced cooling in the bubble phases of nuclear pasta can have an observable effect. Continued monitoring of the Cas A NS temperature over the upcoming decade could place some stringent constraints on that physics.

In the preparation of this manuscript the authors became aware of the preliminary results of a similar study (\url{http://www.nucl.phys.tohoku.ac.jp/nusym13/proc/nusym13_Yeunhwan_Lim.pdf}) constraining the symmetry energy using Cas A temperature measurements, which are in broad agreement with our own (without the use of cooling mechanisms in the bubble phases of pasta).

\section{Acknowledgments}
We thank Dany Page for help running NSCool, and Farrukh Fattoyev for helpful discussions.
This work is supported in part by the National Aeronautics and Space Administration under grant NNX11AC41G issued
through the Science Mission Directorate, the National Science
Foundation under Grants No. PHY-0757839, No. PHY-1068022, US Department of Energy Grants DE-FG02-08ER41533, desc0004971 and the REU program under grant no. PHY-1062613.

\bibliographystyle{apj}

\clearpage


\begin{figure}
\resizebox{1.0\textwidth}{!}{\includegraphics{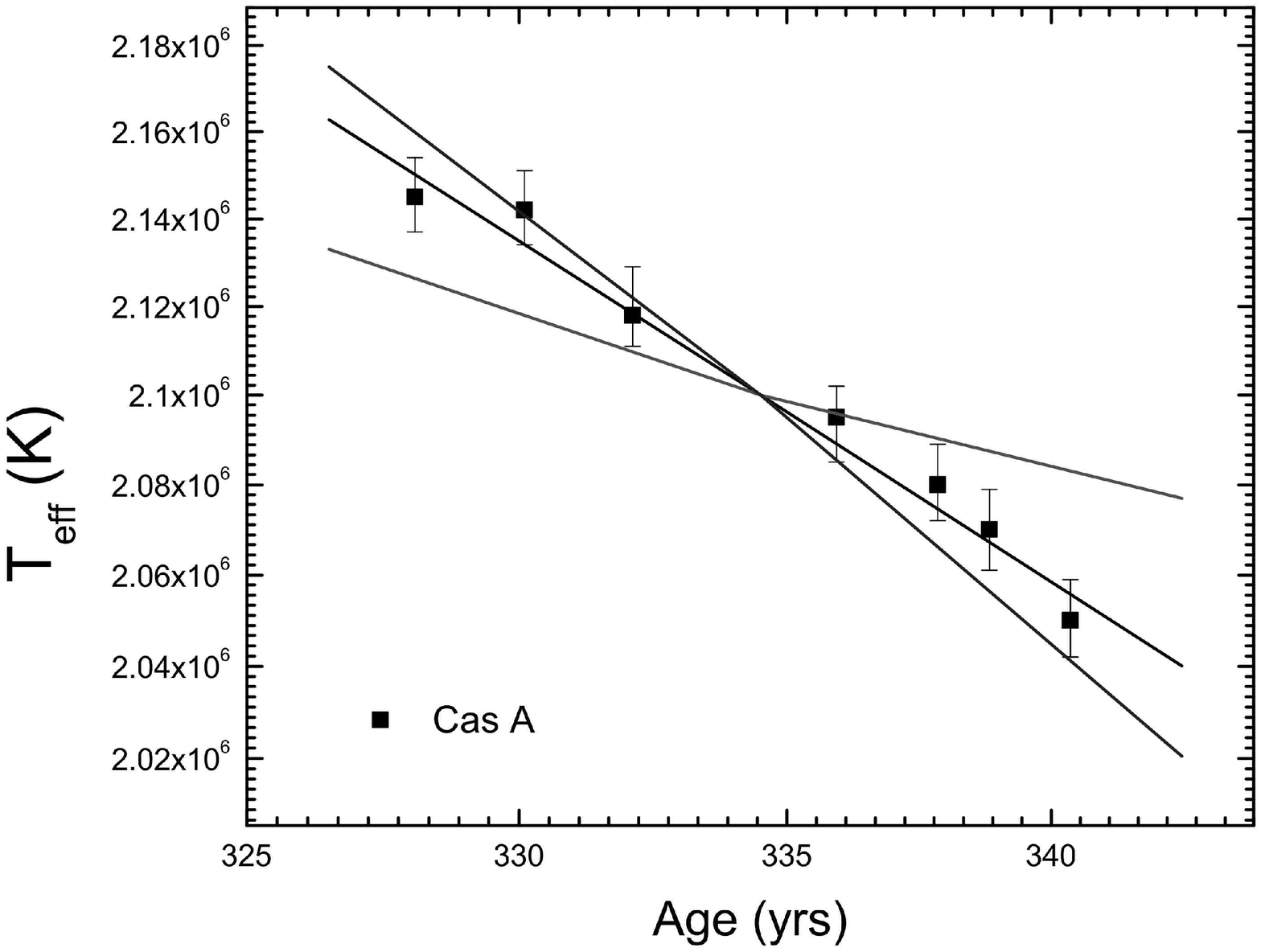}}
\caption{(Color online). Temperature measurements of the Cas A neutron star over the period 2000-2009 using ACIS-S graded observations (data points). Lines show their best fit, upper ($\approx$5.5\% decline) and lower ($\approx$2\% decline) limits on the cooling rate when data from all other \emph{Chandra} detectors and modes are included \citep{Elshamouty2013}.}\label{Fig1}
\end{figure}

\clearpage


\begin{figure}
\includegraphics{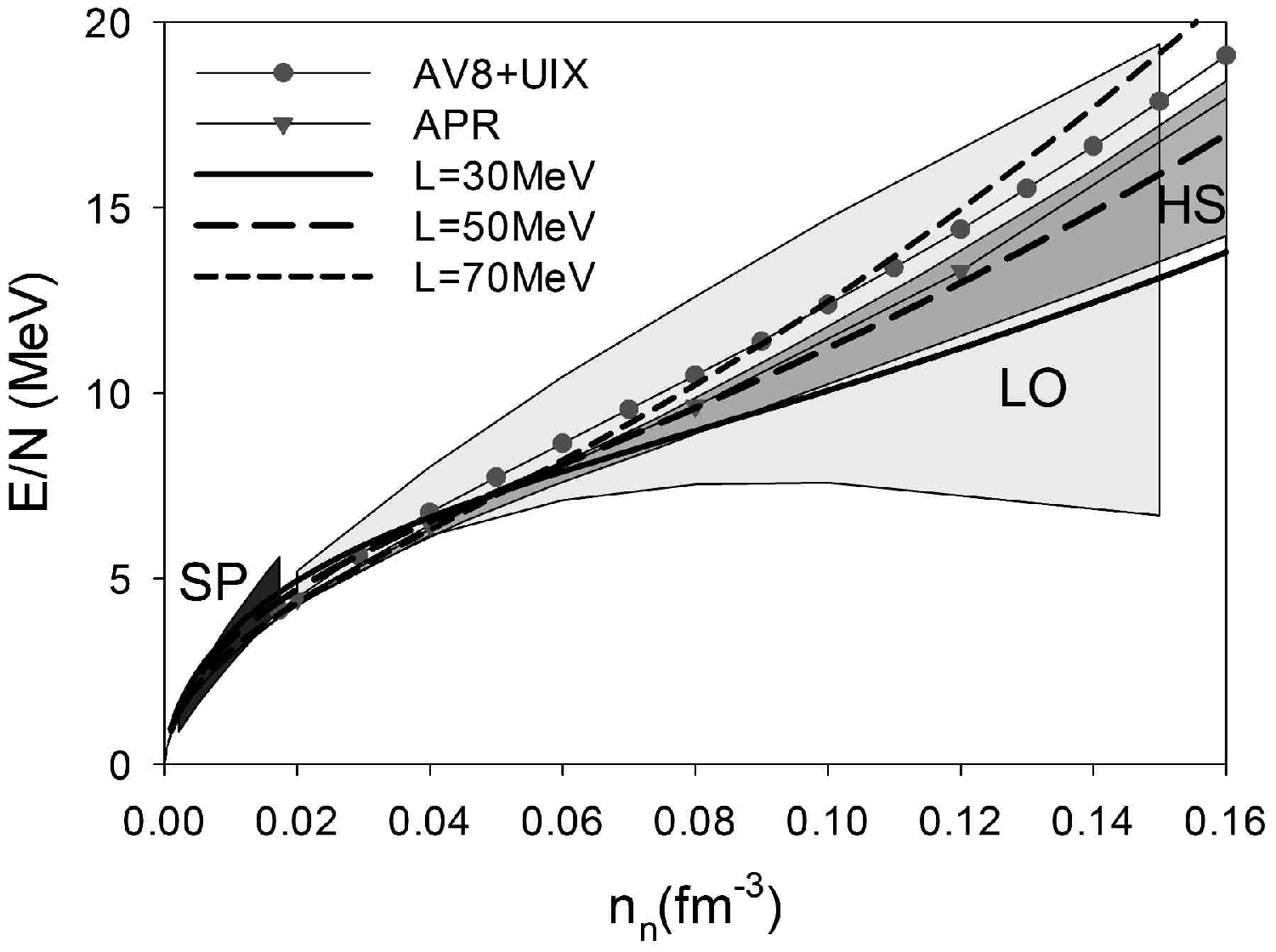}
\caption{(Color online). Energy per neutron versus neutron baryon density for pure neutron matter obtained from calculations of Fermi gases in the unitary limit \citep{Schwenk2005} (SP), chiral effective field theory \citep{Hebeler2010} (HS), quantum Monte Carlo calculations using chiral forces at leading order \citep{Gezerlis2013} (LO),  Auxiliary Field Diffusion Monte Carlo using realistic two-nucleon interactions plus phenomenological three-nucleon interactions AV8+UIX \citep{Gandolfi2009,Gandolfi2010}, and the APR EOS \citep{Akmal1998}. Results using the Skyrme model SkIUFSU used in this paper are shown for $L$=30, 50 and 70 MeV}\label{Fig2}
\end{figure}

\clearpage


\begin{figure}[t!]
\begin{center}
\resizebox{0.55\textwidth}{!}{\includegraphics{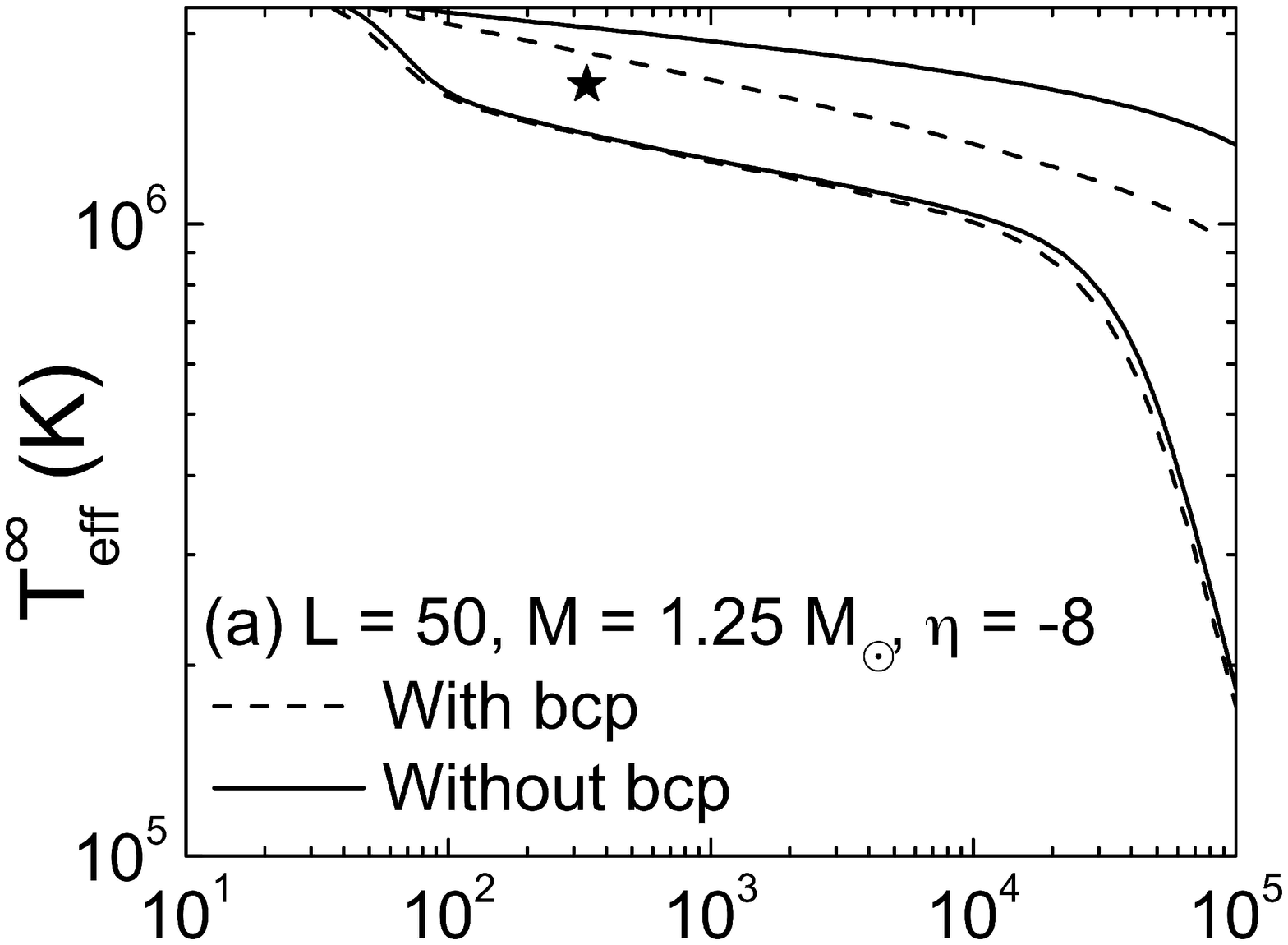}{\includegraphics{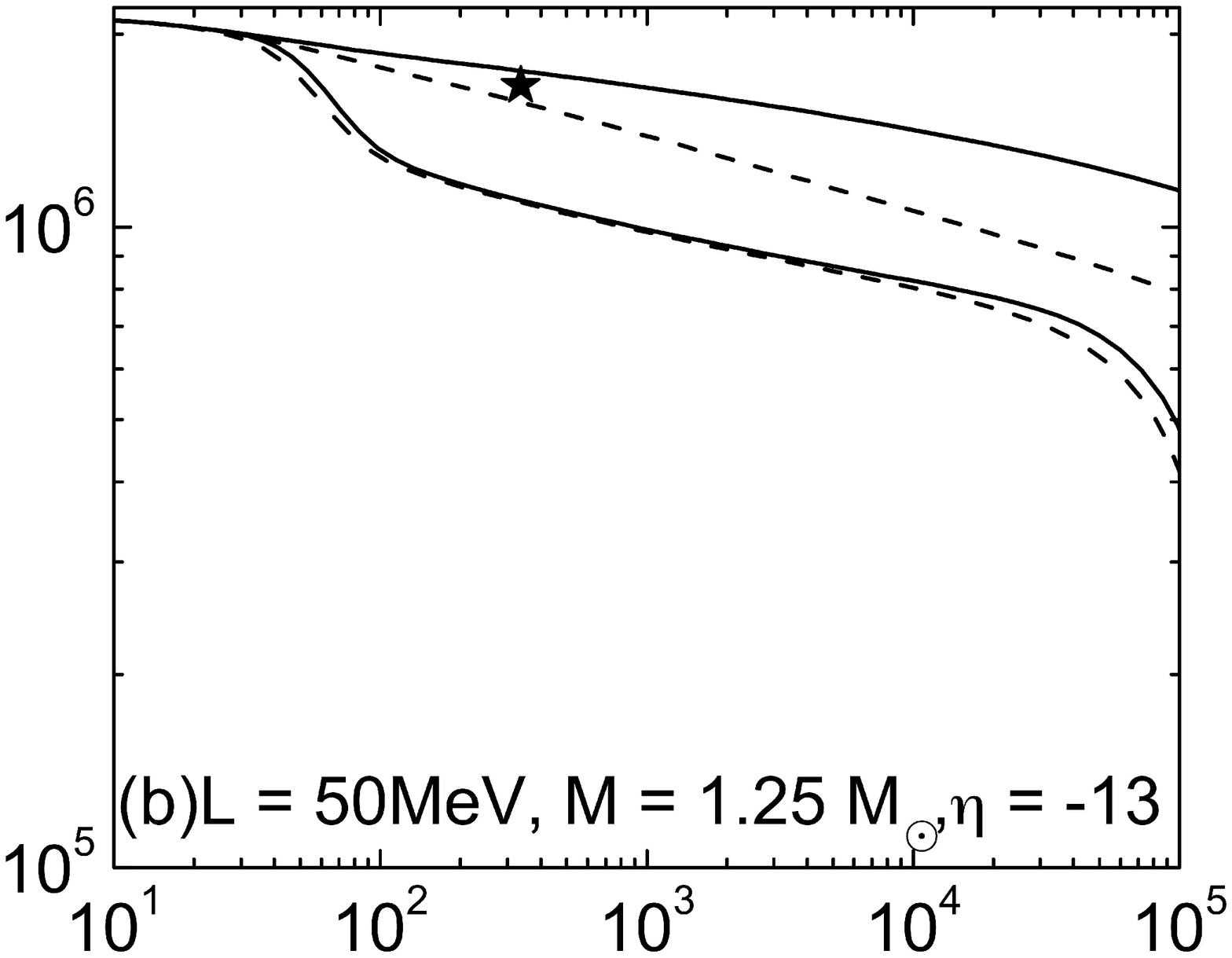}}}
\resizebox{0.55\textwidth}{!}{\includegraphics{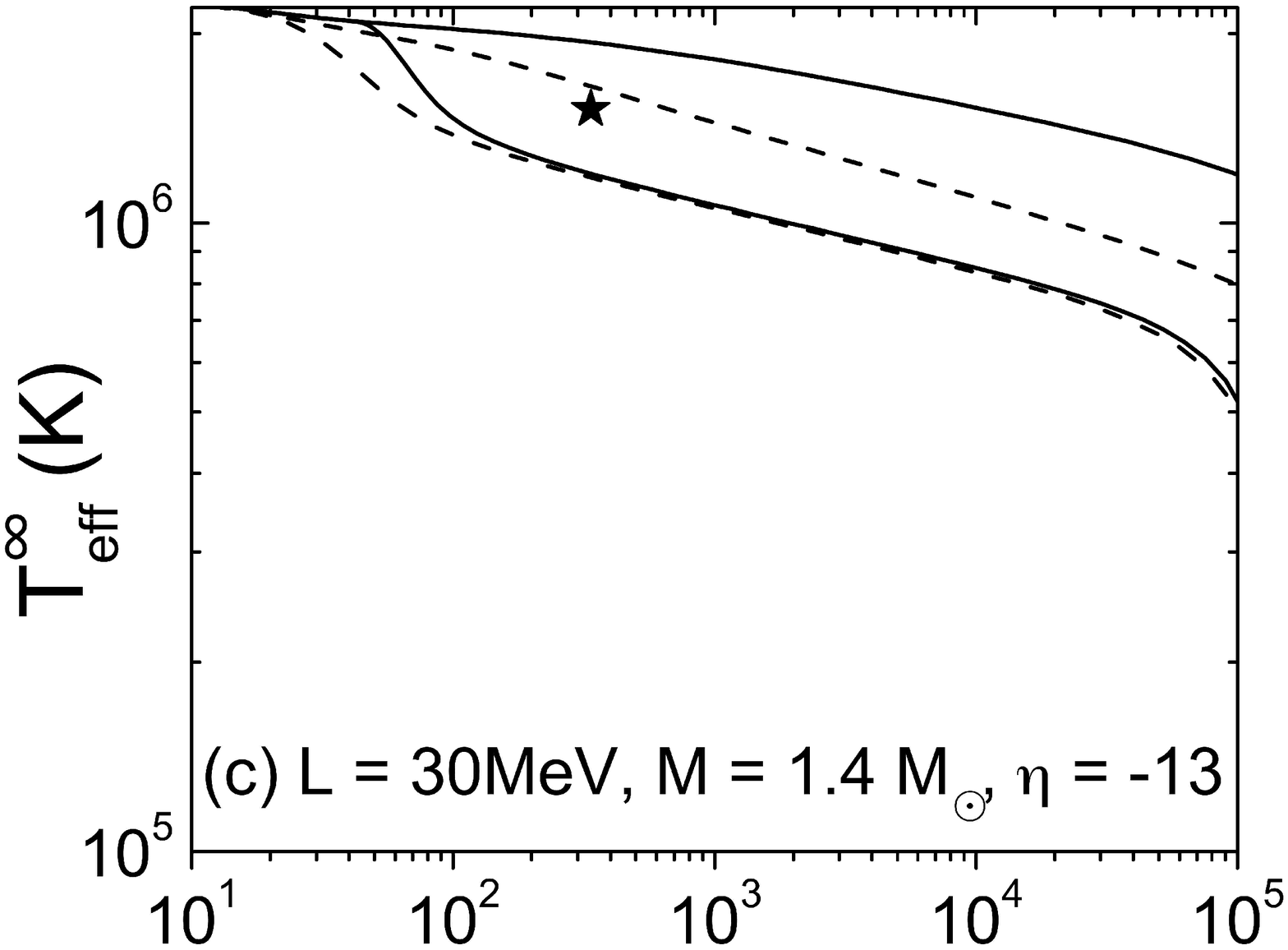}{\includegraphics{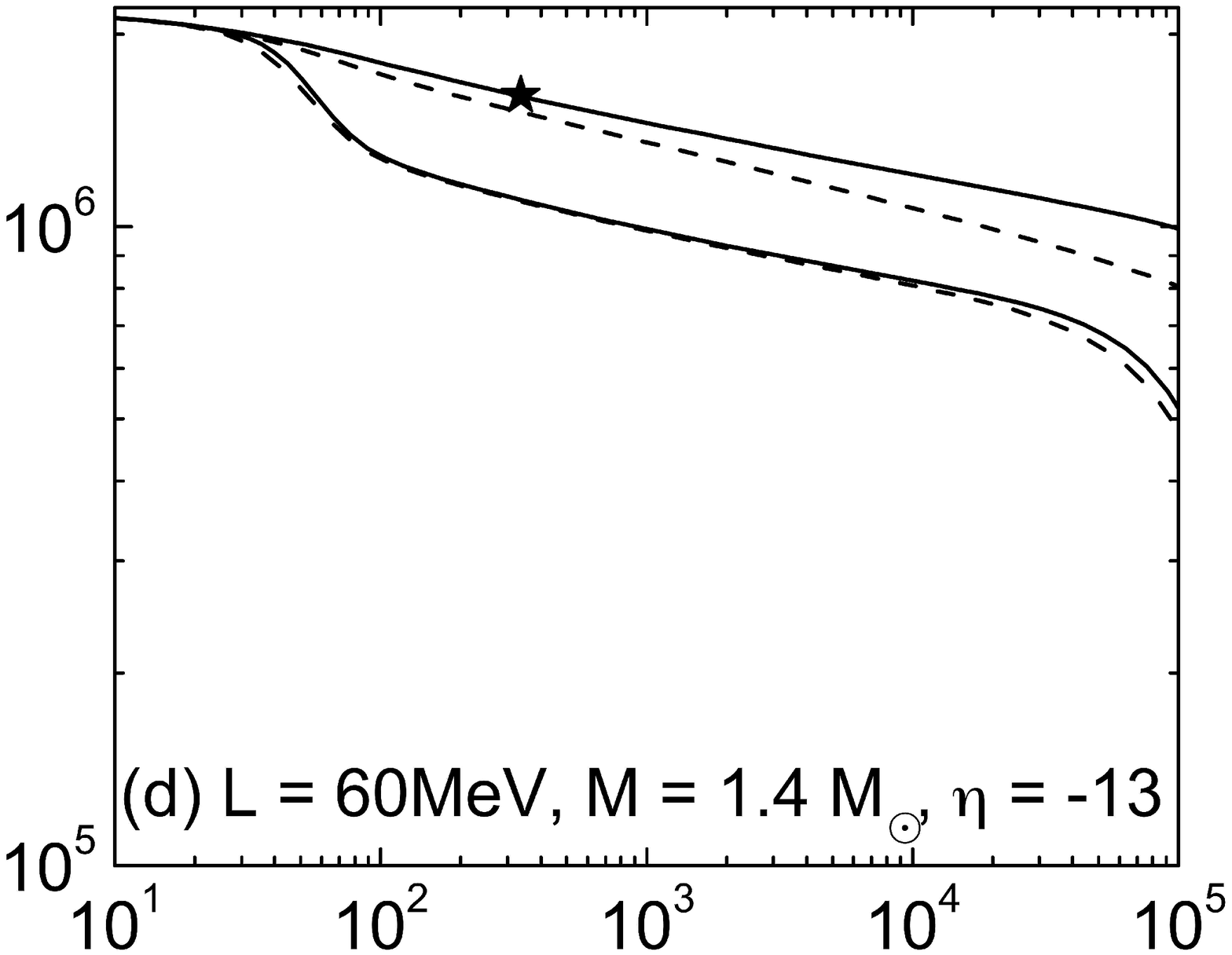}}}
\resizebox{0.55\textwidth}{!}{\includegraphics{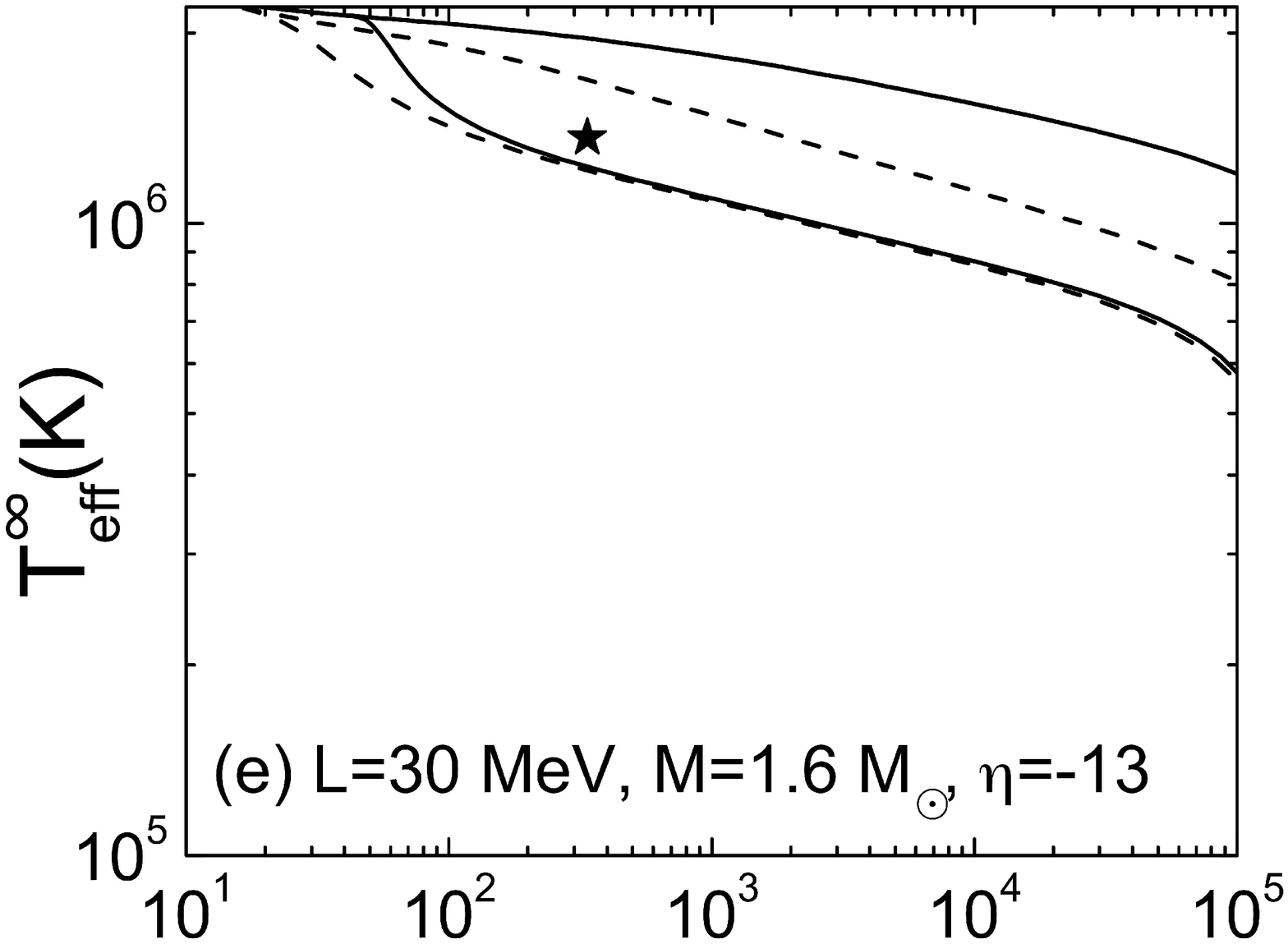}{\includegraphics{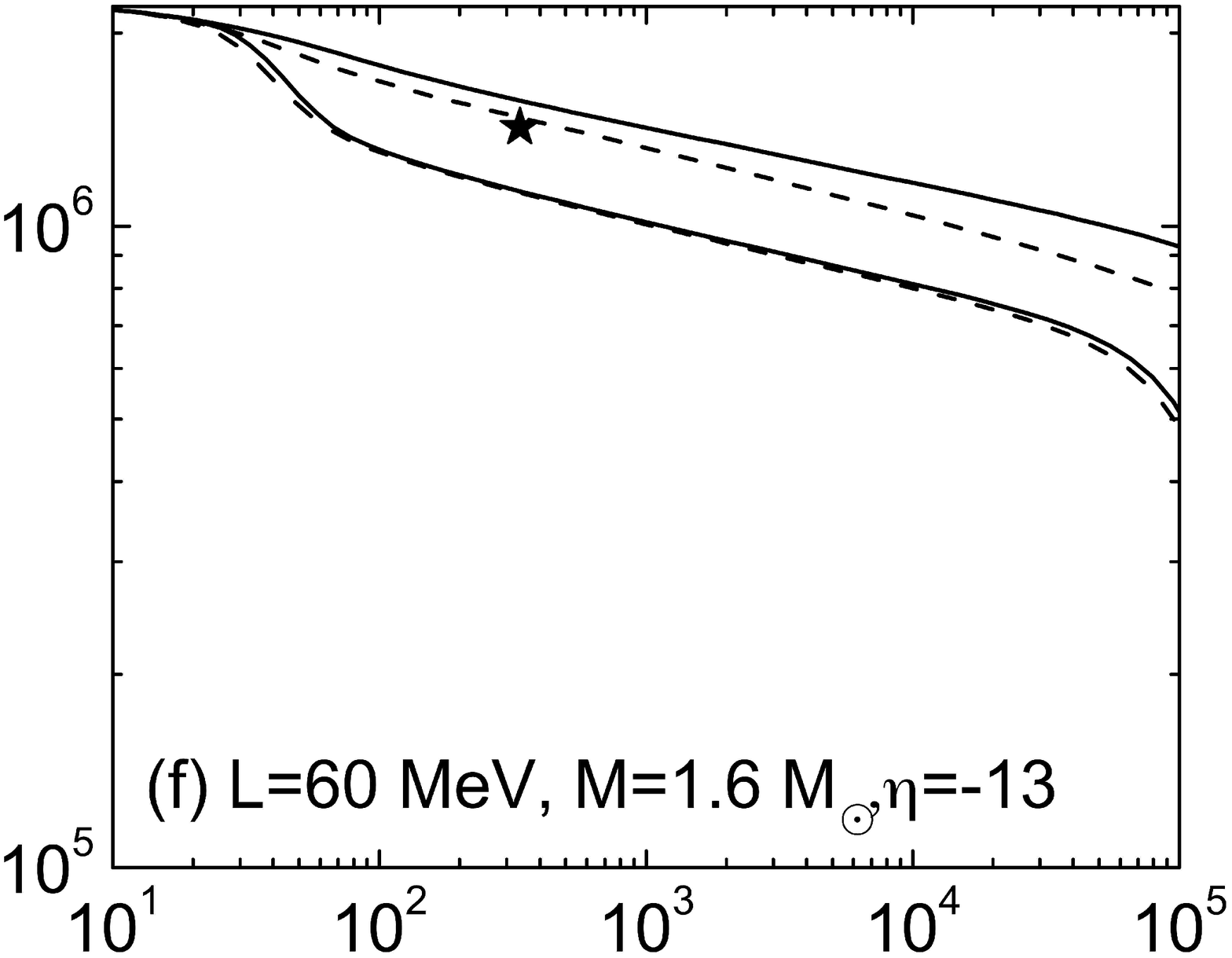}}}
\resizebox{0.55\textwidth}{!}{\includegraphics{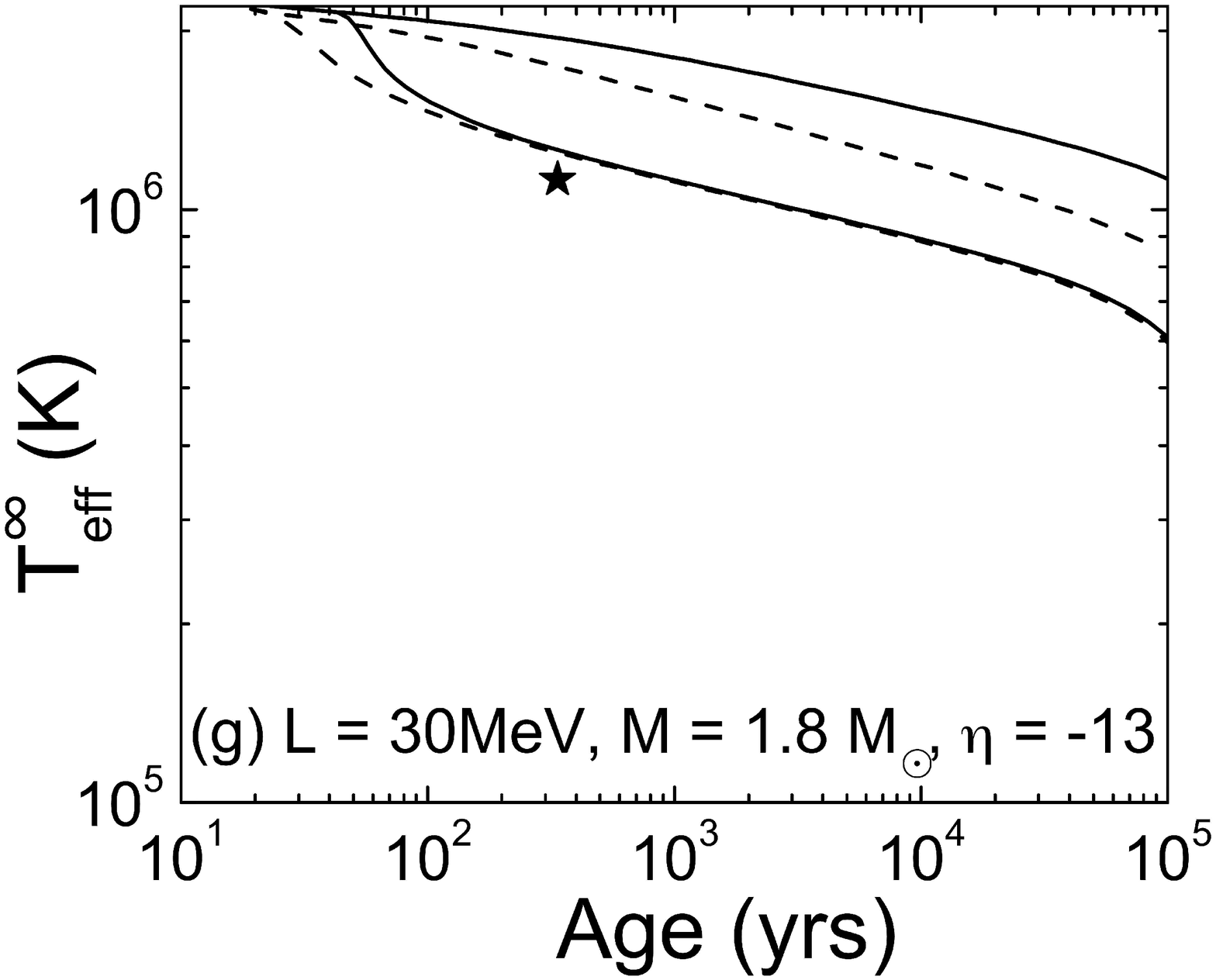}{\includegraphics{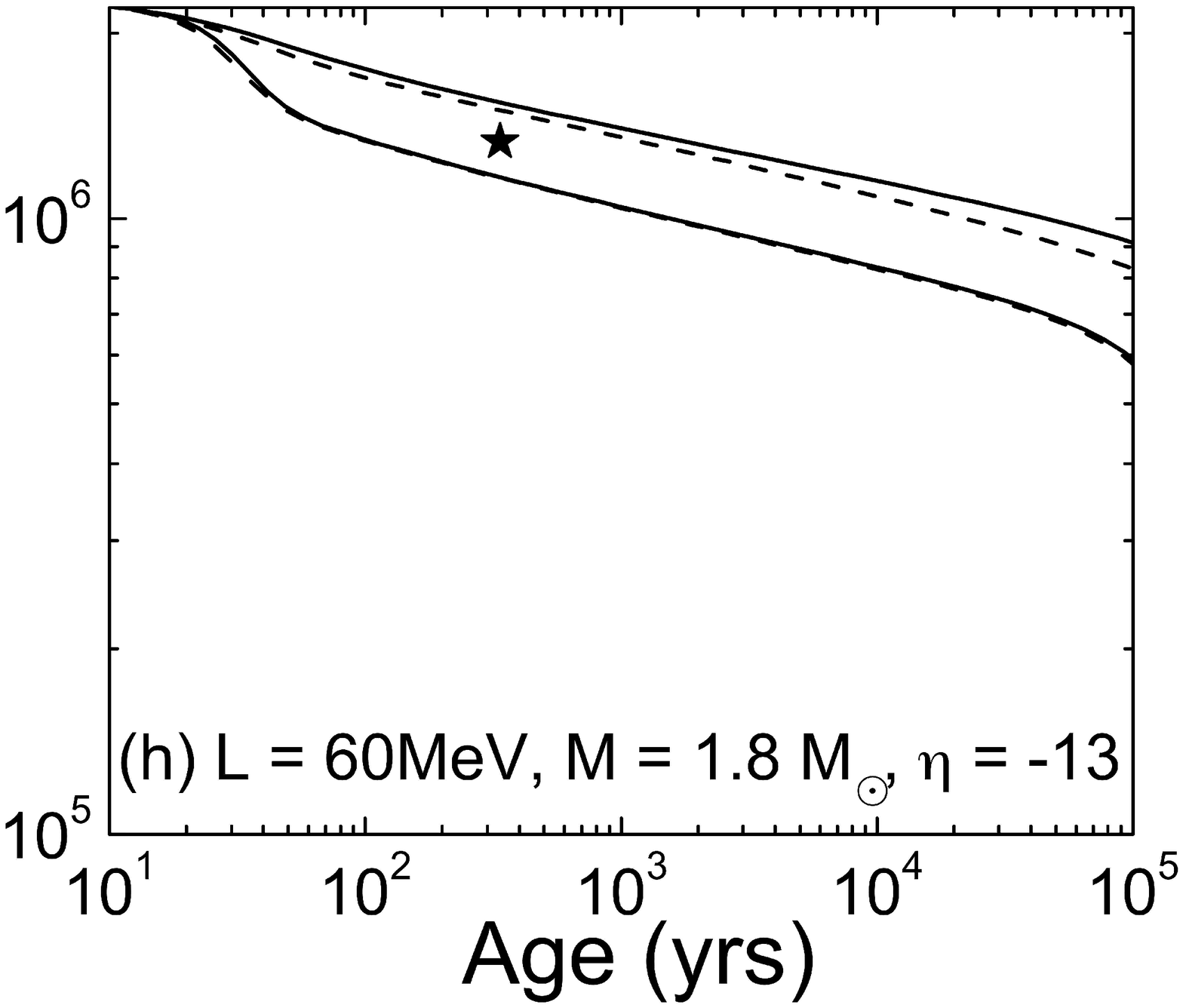}}}
\end{center}
\caption{Comparison of cooling curves $T_{\rm eff}^{\infty}(t)$ with the \emph{average} temperature of the CANS under variations of model parameters. The measured surface temperature is appropriately redshifted for the given stellar model ($M,R$). Each plot shows two pairs of cooling curves; one with BCPs active (dashed) and one with BCPs inactive (solid). In each pair, the upper curve corresponds to $T_{\rm cn}^{\rm max}$ = 0K (no core neutron superfluidity) and the lower to $T_{\rm cn}^{\rm max}$ = $10^9$K, defining the cooling window within which the CANS temperature should lie. Figs 3a and b illustrate the effect of changing the envelope composition from $\eta=-8$ to $\eta=-13$ respectively with $L$=50 MeV and $M=1.25M_{\odot}$; Figs 3c-f illustrate the effect of changing $L$ from 30 to 60 MeV (c-d and e-f respectively) and mass $M$ from $M=1.4M_{\odot}$ to $M=1.8M_{\odot}$ (c-e-g and d-f-h respectively) with $\eta=-13$.}\label{Fig3}
\end{figure}

\clearpage


\begin{figure*}[t!]
\resizebox{1.0\textwidth}{!}{\includegraphics{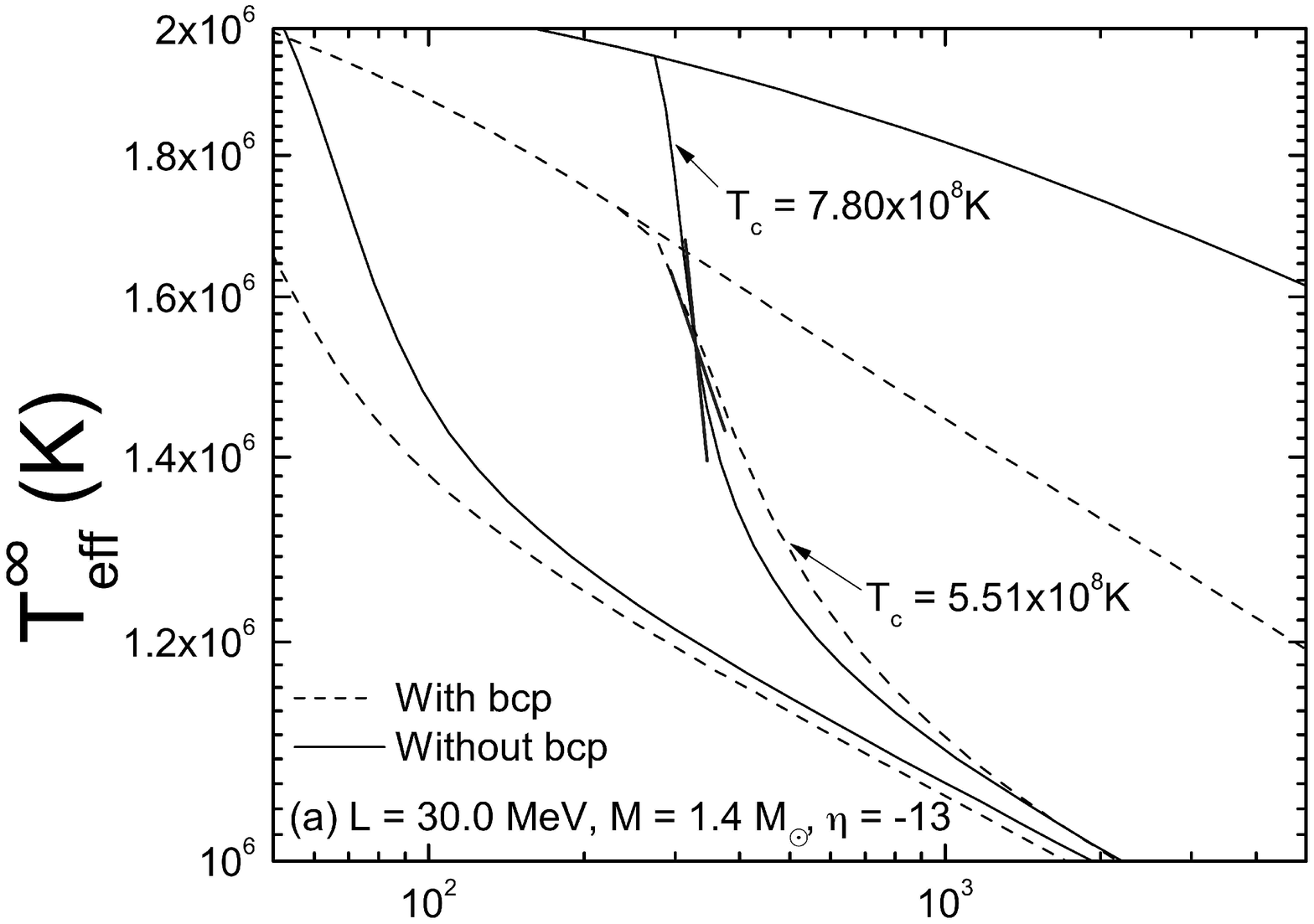}{\includegraphics{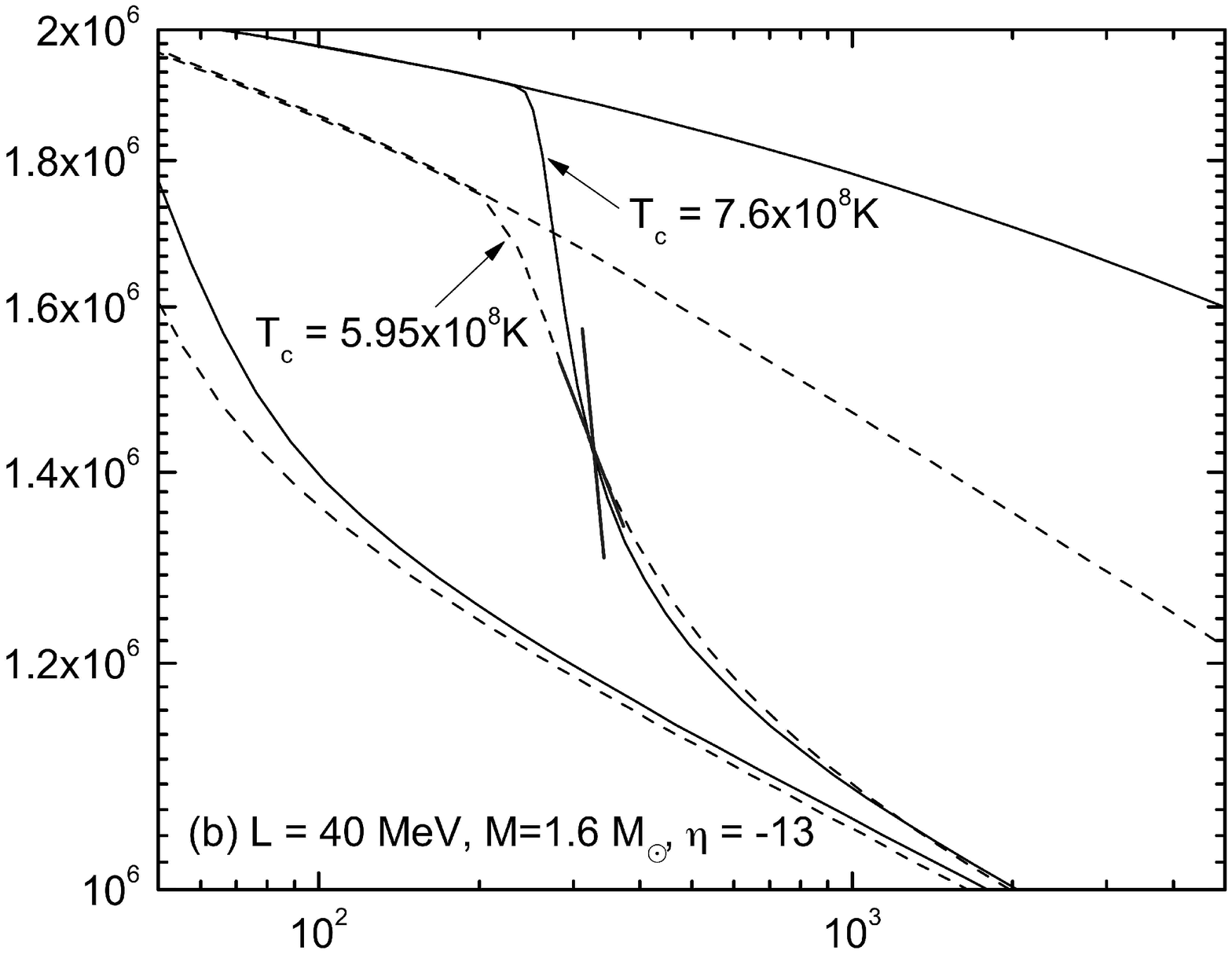}}}
\resizebox{1.0\textwidth}{!}{\includegraphics{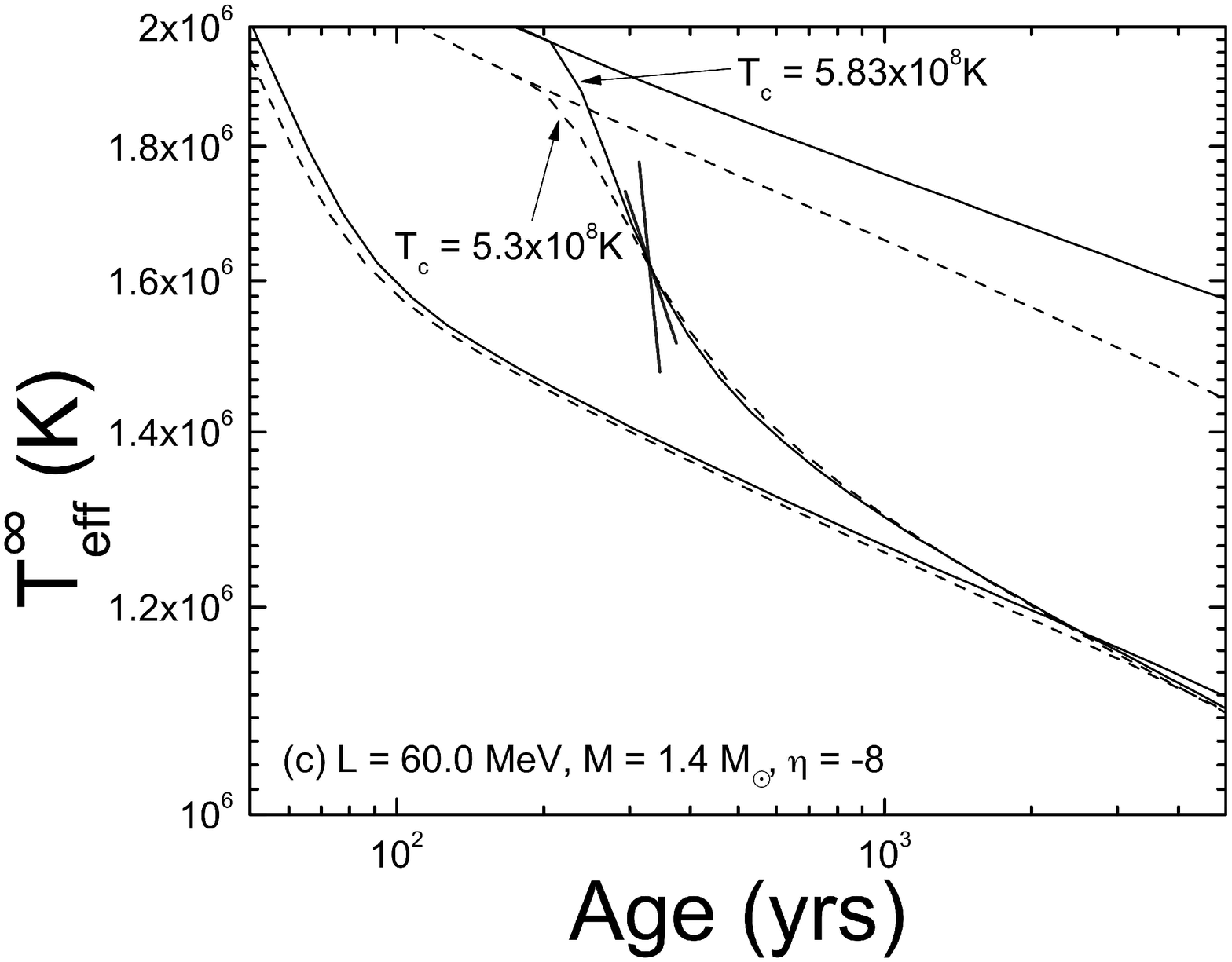}{\includegraphics{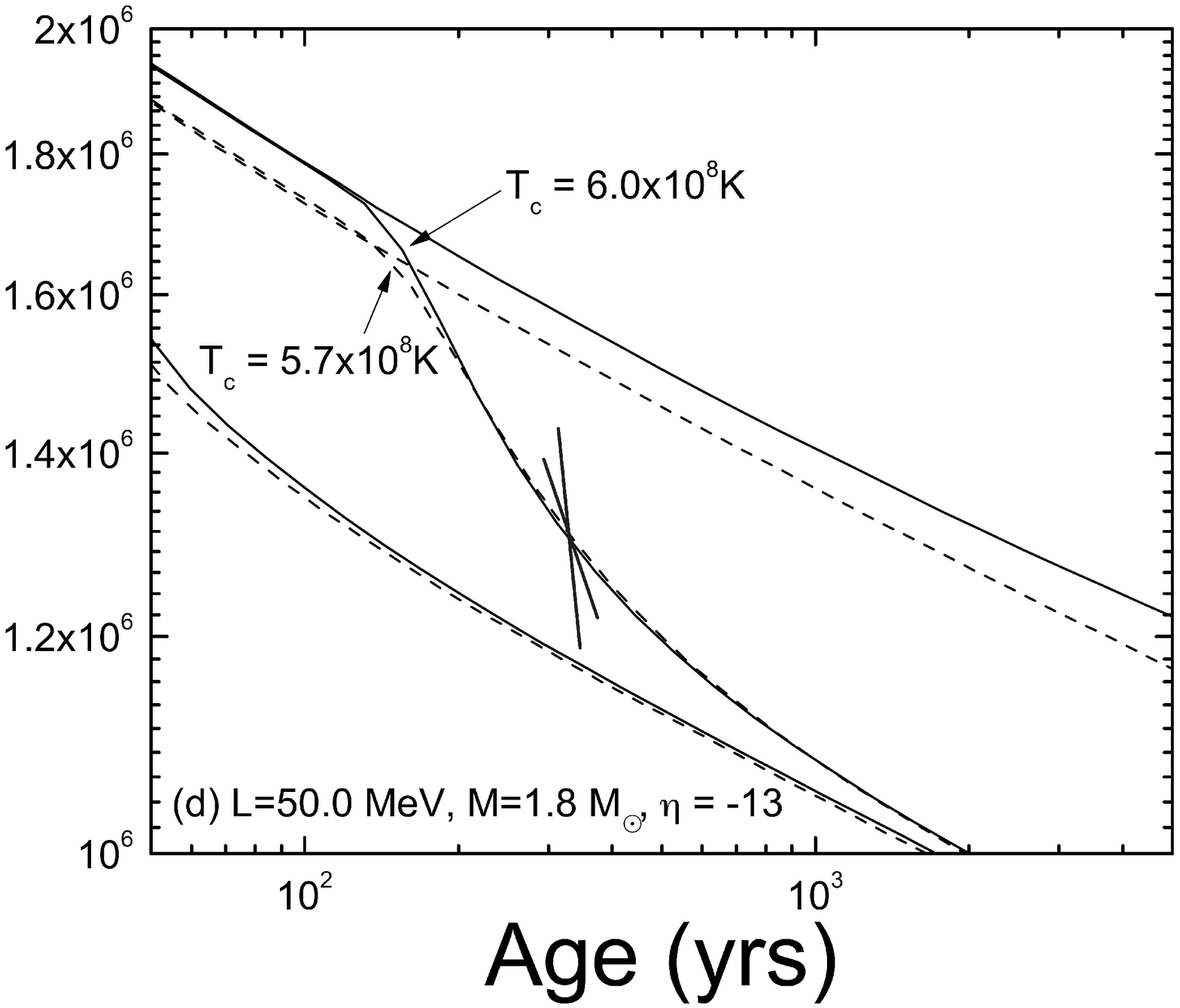}}}
\caption{4 pairs of cooling windows, together with the cooling curves which pass through the average value of the measured CANS temperature and the corresponding value of $T_{\rm cn}^{\rm max}$. Each plot shows the window with BCPs inactive (solid lines) and active (dashed lines) for combinations ($L$(MeV), $M/M_{\odot}$, $\eta$) of (30, 1.4, -13) (a), (40, 1.6, -13) (b), (60, 1.4, -8) (c) and (50, 1.8, -13) (d). The steepest and shallowest temperature declines estimated from observations are indicated by the straight lines passing through the average CANS temperature. The inferred rapid cooling prefers lower values of $L$, $M$ or $\eta$.}\label{Fig4}
\end{figure*}

\clearpage


\begin{table*}
\begin{center}
\caption{Ranges of the slope of the symmetry energy $L$(MeV) whose cooling curves pass through the average observed temperature of the Cas A NS only (top), and ranges whose cooling curves additionally fall within the limits of the observed cooling rate (bottom) for a conservative range of masses $M$, envelope compositions $\eta$, and with and without BCPs. A dashed line indicates no matching cooling curves were found for that particular parameter combination. \label{Tab1}}
\begin{tabular}{ccccc}
\tableline
$M(M_{\odot}$) & $\eta$=-8; BCP & $\eta$=-13; BCP & $\eta$=-8; no BCP & $\eta$=-13; no BCP \\

\tableline

1.25 & $\lesssim$ 70 & - & $\lesssim$ 70 & $\lesssim$ 55 \\
1.40 & $\approx$ 35-65 & $\lesssim$ 45 & $\lesssim$ 65 &  $\lesssim$ 55 \\
1.60 & $\approx$ 55-65 & $\lesssim$ 55 & $\lesssim$ 55-65 &  $\lesssim$ 65 \\
1.80 & - & $\approx$ 45-65 & - &  $\approx$ 45-65 \\

\tableline

1.25 & $\lesssim$ 45 & - & $\lesssim$ 70 & $\lesssim$ 55 \\
1.40 & - & $\lesssim$ 35 & $\lesssim$ 55 &  $\lesssim$ 55 \\
1.60 & - & $\approx$ 35-45 & - &  $\approx$ 35-55 \\
1.80 & - & - & - &  - \\

\tableline

\end{tabular}

\end{center}
\end{table*}

\end{document}